\documentstyle[12pt]{article}
\def\hybrid{\topmargin 0pt      \oddsidemargin 0pt
	\headheight 0pt \headsep 0pt
	\textheight 9in         % US paper
	\textwidth 6.25in       % A4 paper
	\marginparwidth .875in
	\parskip 5pt plus 1pt   \jot = 1.5ex}

\catcode`\@=11
\def\marginnote#1{}
\newcount\hour
\newcount\minute
\newtoks\amorpm
\hour=\time\divide\hour by60
\minute=\time{\multiply\hour by60 \global\advance\minute by-\hour}
\edef\standardtime{{\ifnum\hour<12 \global\amorpm={am}%
	\else\global\amorpm={pm}\advance\hour by-12 \fi
	\ifnum\hour=0 \hour=12 \fi
	\number\hour:\ifnum\minute<10 0\fi\number\minute\the\amorpm}}
\edef\militarytime{\number\hour:\ifnum\minute<10 0\fi\number\minute}

\def\draftlabel#1{{\@bsphack\if@filesw {\let\thepage\relax
   \xdef\@gtempa{\write\@auxout{\string
      \newlabel{#1}{{\@currentlabel}{\thepage}}}}}\@gtempa
   \if@nobreak \ifvmode\nobreak\fi\fi\fi\@esphack}
	\gdef\@eqnlabel{#1}}
\def\@eqnlabel{}
\def\@vacuum{}
\def\draftmarginnote#1{\marginpar{\raggedright\scriptsize\tt#1}}

\def\draft{\oddsidemargin -.5truein
	\def\@oddfoot{\sl preliminary draft \hfil
	\rm\thepage\hfil\sl\today\quad\militarytime}
	\let\@evenfoot\@oddfoot \overfullrule 3pt
	\let\label=\draftlabel
	\let\marginnote=\draftmarginnote
   \def\@eqnnum{(\theequation)\rlap{\kern\marginparsep\tt\@eqnlabel}%
\global\let\@eqnlabel\@vacuum}  }

\def\numberbysection{\@addtoreset{equation}{section}
	\def\theequation{\thesection.\arabic{equation}}}

\def\underline#1{\relax\ifmmode\@@underline#1\else
	$\@@underline{\hbox{#1}}$\relax\fi}

\def\titlepage{\@restonecolfalse\if@twocolumn\@restonecoltrue\onecolumn
     \else \newpage \fi \thispagestyle{empty}\c@page\z@
	\def\thefootnote{\fnsymbol{footnote}} }

\def\endtitlepage{\if@restonecol\twocolumn \else  \fi
	\def\thefootnote{\arabic{footnote}}
	\setcounter{footnote}{0}}  %\c@footnote\z@ }
\catcode`@=12
\relax
\def\ie{\hbox{\it i.e.}}        
        
\def\beq{\begin{equation}}
\def\eeq{\end{equation}}
\def\bea{\begin{eqnarray}}
\def\eea{\end{eqnarray}}
\def\nn{\nonumber}
\def\ep{\epsilon}

\relax
\def\half{{1\over 2}\;}

\hybrid
\begin{document}
\begin{titlepage}
\begin{center}
April~1994 \hfill    PAR--LPTHE 94/17 \\[.5in]
{\large\bf Spin--spin critical point correlation
functions for the 2D random bond Ising and Potts models}\\[.5in]
        \bf     Vladimir Dotsenko\footnote{Also at the Landau Institute for
Theoretical Physics, Moscow}, Marco Picco and Pierre
Pujol \\
	{\it LPTHE\/}\footnote{Laboratoire associ\'e No. 280 au CNRS}\\
       \it  Universit\'e Pierre et Marie Curie, PARIS VI\\
	Boite 126, Tour 16, 1$^{\it er}$ \'etage \\
	4 place Jussieu\\
	F-75252 Paris CEDEX 05, FRANCE\\
	dotsenko@lpthe.jussieu.fr, picco@lpthe.jussieu.fr,
pujol@lpthe.jussieu.fr
\end{center}

\vskip .5in
\centerline{\bf ABSTRACT}
\begin{quotation}
We compute the combined two and three loop order correction to the spin-spin
correlation functions for the 2D Ising and q-states Potts model with random
bonds at the critical point. The procedure employed is the renormalisation
group approach for the perturbation
series around the conformal field theories representing the pure models.
We obtain corrections for the correlations functions which
produce crossover in the amplitude but don't change
the critical exponent in the case of the Ising model and which produce
a shift in the critical exponent,
due to randomness, in the case of the Potts model. Comparison with
numerical data is discussed briefly.

\end{quotation}
\end{titlepage}
\newpage

In severals works, the critical properties of the two dimensional Ising and
Potts models with quenched random bonds has been investigated in a
numerical and theoretical way. For the Ising model, it seems that the
effect of impurities changes the divergent form of the specific heat near
the critical point $C(t) \sim lnln\left(1 \over |t| \right)$ \cite{dots1}
while the critical exponent of the $<\sigma \sigma>$ correlation function
is unchanged \cite{shalaev,ludwig2}. The renormalisation group (R.G)
equations for the coupling constant as well as the correction for the
$<\varepsilon \varepsilon>$ correlation function has been established to
the second loop order \cite{ludwig1}.  Recent numerical data gives
interesting behavior in the crossover region for the $<\sigma \sigma>$
correlation function
\cite{talapov}. These results suggest a deviation of such a correlation
function in comparison to the pure case, deviation which is not predicted
in previous perturbative computation up to the first order \cite{ludwig2}.
We will extend in this letter the computation of this correlation function
up to the third order. We found that there is a multiplicative correction
in $<\sigma \sigma>$ (compared to the pure $2d$ Ising model). We then
compare these corrections with the ones of \cite{talapov}.  We also
consider the random Potts model, for which we introduce a perturbed central
charge ($c={1\over 2} + \epsilon$), which can be seen as a short distance
regulator\footnote{Such a procedure was already used by A.~Ludwig in the
computation of the thermal exponent \cite{ludwig1}.}.  We obtain
the correction for the critical exponent of the $<\sigma \sigma>$ up to the
third order. The choice of the conformal field theory which should
represent exactly (in a non perturbative way) the Potts model with random
bonds at the critical point remains open and a complete non perturbative
characterization of the critical exponents is needed.

Our starting point is the Hamiltonian at the critical temperature :
\beq
S_0 =\displaystyle\sum_{<i,j>} (-\beta J_0) \delta_{s_i,s_j}
\eeq
where $J_0$ is the coupling between bonds and $s_i$ are the Ising or Potts
spin variables. The unperturbed partition function is obtained by taking
a sum over the spins configurations~:
\beq
Z_0 = \sum_{s_i} e^{-S_0}
\eeq
Near the critical point, the addition of a position dependent random
coupling constant gives the following effective Hamiltonian expressed in a
continues way :
$$
S =  S_0 + \int d^2z\; m(z) \varepsilon(z)
$$
Here $ \varepsilon(z)$ is the local energy operator of the critical pure
model. Using the replica method \cite{dots1,ludwig1}, one can find
the free energy by averaging over a distribution of the random function
$m(z)$ :
$$
\overline{\log(Z)} = \displaystyle\lim_{N\to0}{\overline{{Z^N - 1 \over N}}}
$$
with :
\beq
Z^N =  \sum_{s_i}
e^{-\displaystyle\sum_{a=1}^N S_{0,a} - \int d^2z\;
m(z)\displaystyle\sum_{a=1}^N \varepsilon_a(z) }
\eeq
The average of $ Z^N$ with an arbitrary distribution for $m(z)$ gives an
effective Hamiltonian containing all the cumulants of the probability
measure. However, it can be shown by power counting that for the q-state
Potts model all but the two firsts cumulants are irrelevant for $q \leq 3$
\cite{ludwig1}. The effective Hamiltonian can then be obtained by averaging
the partition function with a Gaussian distribution :
$$
\overline{Z^N} = \int \prod_z dm(z) Z^N e^{-{1\over2g_0}(m(z)-m_0)^2}
$$
which gives :
\beq
\overline{Z^N} = \sum_{s_i}
e^{-\displaystyle\sum_{a=1}^N S_{0,a}
 + g_0 \int d^2z\; \displaystyle\sum_{a,b=1}^N
\varepsilon_a(z)\varepsilon_b(z)  - m_0 \int d^2z\; \displaystyle\sum_{a=1}^N
\varepsilon_a(z) }
\eeq
$m_0$ represents the deviation from the critical temperature of the model
with random bonds and we are interested in the limit $m_0 \rightarrow 0$.
Inside $\int d^2z\; \displaystyle\sum_{a,b=1}^N
\varepsilon_a(z)\varepsilon_b(z) $,
the terms with coinciding replica number ($a=b$) can be omitted because
they produce, under an operator product expansion (OPE), only divergent
constant terms which can be removed by subtraction and irrelevant
operators.  Diagonal replica number terms will then be discarded in the
following and the partition function can be written like
\beq
\sum_{s_i} e^{-\displaystyle\sum_{a=1}^N
S_{0,a}
 + g_0 \int d^2z\; \displaystyle\sum_{a\not=b}^N
\varepsilon_a(z)\varepsilon_b(z)
- m_0 \int d^2z\; \displaystyle\sum_{a=1}^N \varepsilon_a(z) }
\eeq
Correlation functions are calculated in a perturbative way around the
conformal invariant action $S_0$. So a correlation function of some local
operator $O$ is expanded like :
$$
<O(0)O(R)> = <O(0)O(R)>_0+<S_IO(0)O(R)>_0+{1\over2}<S_I^2O(0)O(R)>_0+\cdots
$$
$<>_0$ means an expectation value taken with
respect to $S_0$ and
\beq
S_I =
\int d^2z\; H_I(z)  =  g_0 \int d^2z\; \displaystyle\sum_{a\not=b}^N
\varepsilon_a(z)\varepsilon_b(z)
- m_0 \int d^2z\; \displaystyle\sum_{a=1}^N \varepsilon_a(z)
\eeq
The integrals of correlation functions involved in the calculation can be
performed by analytic continuation with the Coulomb-gas representation of a
conformal field theory \cite{dots2} where the central charge is $c = {1
\over 2} +
\epsilon$. The $\epsilon$ term corresponds to a short distance regulator
for the integrals. In addition, we also used an I.R. cut-off $r$. The
result is then expressed in a $\epsilon$ series with coefficients depending
on $r$. Then the limit $\epsilon \rightarrow 0$ corresponds to the Ising
model while the Potts model is obtained for some finite value of
$\epsilon$. We recall here some notations of the Coulomb-gas representation
for the vertex operators \cite{dots2}. The central charge $c$ is
characterized by $\alpha^2_+ = {2p\over2p-1} ={4\over 3} + \ep $ with
\bea
\label{c}
c=1-24 \alpha_0^2 \hskip 0.5cm &;& \hskip 0.5cm \alpha_\pm = \alpha_0 \pm
\sqrt{\alpha_0^2 +1} \\
\alpha_+ \alpha_- &=& -1 \nonumber
\eea
Note that for the pure 2D Ising model
 $\alpha_+^2 ={4\over 3} $ and $c = {1\over 2}$. The vertex operators are
\beq
V_{nm}(x) = e^{i\alpha_{nm} \phi(x)}
\eeq
with $\phi(x)$ a free scalar field and $\alpha_{nm}$ defined by
\beq
\alpha_{nm}=\half (1-n) \alpha_- + \half (1-m) \alpha_+
\eeq
The conformal dimension of an operator $V_{nm}(x)$ is
$\Delta_{nm} =
-\alpha_{\overline{nm}} \alpha_{nm}$ with
\beq
\label{al}
\alpha_{\overline{nm}} = 2\alpha_0 -\alpha_{nm} =\half (1+n) \alpha_- +
\half (1+m) \alpha_+
\eeq
The spin field can be represented by the vertex operator
$V_{p,p-1}$ and $V_{1,2}$ for the energy operator $\varepsilon$.
Note that in the Ising case the $\sigma$ operator could also be represented
by  the $V_{21}$ operator (since both operators coincide in the limit
$\epsilon \rightarrow 0$) and the final result turn out to be
independent on which representation we are taking \cite{dpp}.

We now turn to the computation of
the renormalisation of $\sigma$ produced by the  operator
product
\beq
{g_0^k\over k!}  \prod_{i}^{k} \int \limits_{|x_i-z|<r} d^2x_i d^2z\;
\displaystyle\sum_{a_i\not= b_i}^N
\varepsilon_{a_i}(x_i) \varepsilon_{b_i}(x_i)   \displaystyle\sum_{e=1}^N
\sigma_e(z)   \rightarrow    g_0^k
X(r,\epsilon)\int d^2z  \displaystyle\sum_{a=1}^N \sigma_a(z)
\eeq
The asymptotic behavior of the
correlation function can be calculated with the help of the R.G. equations.
A straightforward calculation yields a null contribution at the first
order. The next two orders are
\beq
\label{deux}
{g_0^2\over2}  \int\limits_{|x-z|,|y-z|<r} d^2x d^2y d^2z
\displaystyle\sum_{a\not= b}^N
\varepsilon_a(x) \varepsilon_b(x)  \displaystyle\sum_{c\not= d}^N
\varepsilon_c(y) \varepsilon_d(y) \displaystyle\sum_{e=1}^N
\sigma_e(z)
\eeq
and
\beq
\label{trois}
{g_0^3\over3!} \int\limits_{|x-z|,|y-z|,|w-z| <r} d^2x d^2y d^2w d^2z
\displaystyle\sum_{a\not= b}
\varepsilon_a(x) \varepsilon_b(x)  \displaystyle\sum_{c\not= d}
\varepsilon_c(y) \varepsilon_d(y)
 \displaystyle\sum_{e\not= f}\varepsilon_e(w) \varepsilon_f(w)
\displaystyle\sum_{g}\sigma_g(z)
\eeq
The renormalisation constants appearing at each order can be calculated
by projecting (\ref{deux}) and (\ref{trois}) over $\sigma(\infty)$ giving
respectively the following integrals :
\beq
\label{deuxx}
 2 g_0^2 (N-1)  \int\limits_{|x-z|,|y-z|<r} d^2x d^2y
<\sigma(z) \varepsilon(x) \varepsilon(y) \sigma(\infty)> <
\varepsilon(x)  \varepsilon(y)>
\eeq
\bea
\label{troiss}
4(N-1)(N-2) g_0^3   \int\limits_{|x-z|,|y-z|,|w-z| <r} d^2x d^2y d^2w
<\sigma(z) \varepsilon(x) \varepsilon(y) \sigma(\infty)> \times \nonumber \\
 <\varepsilon(x) \varepsilon(w)> <\varepsilon(y)  \varepsilon(w)>
\eea
Using now the Coulomb-Gas formalism, the second order integral (\ref{deuxx})
can be rewritten like :
\bea
2& g_0^2& (N-1) N_2 \int\limits_{|x|,|y|<r} d^2x d^2y \int d^2u <
V_{\overline{p,p-1}}(0)
V_{12}(x) V_{12}(y) V_{p,p-1}(\infty) J_+(u)>{|x
-y|^{-2\Delta_\varepsilon}} \nn \\
&=& 2 g_0^2 (N-1) N_2
\int\limits_{|x|,|y|<r} d^2x d^2y \int d^2u \;
|x|^{2a}|y|^{2a}|u|^{2b}
|x-u|^{2c}|y-u|^{2c}|x-y|^{-2\Delta_\varepsilon}
\eea
with $a=2 (2\alpha_0 - \alpha_{p,p-1}) \alpha_{12}, b=2 (2\alpha_0 -
\alpha_{p,p-1}) \alpha_+, c=2 \alpha_{12} \alpha_+$ and $d=2 \alpha_{12}
\alpha_{12}$.
Here $J_+(u)$ is the screening charge operator
\beq
J_+(u) = e^{i\alpha_+ \phi(u)}
\eeq
and $N_2$ is a normalization coefficient involved in replacing the
correlation function $<~\sigma\varepsilon\varepsilon\sigma>$ by its
Coulomb-Gas representation. At the third order, we have :
\newpage
\bea
&& \hskip 5cm 4(N-1)(N-2) g_0^3  N_2  \times \nn \\
&&\hskip -1cm\int\limits_{|x|,|y|,|w|<r} d^2y  d^2z  d^2w \int d^2u <
V_{\overline{p,p-1}}(0)
V_{12}(y) V_{12}(z) V_{p,p-1}(\infty) J_+(u)>
 {|y-w|^{-2\Delta_\varepsilon }}{|z -w|^{-2\Delta_\varepsilon }} \nn \\
&&\hskip 5cm = 4(N-1)(N-2) g_0^3 N_2 \times \\
&&\hskip -1cm \int\limits_{|x|,|y|,|w|<r} d^2y d^2z d^2w \int d^2u
|y|^{2a}|z|^{2a}|u|^{2b}
|y-u|^{2c}|z-u|^{2c}|y-z|^{2d}
 |y-w|^{-2\Delta_\varepsilon }
|z-w|^{-2\Delta_\varepsilon } \nn
\eea
The details of the calculation of these integrals (and for the
normalization constants $N_2$) will be presented
elsewhere \cite{dpp}. For the calculation of $N_2$, see also \cite{volodya}.
The final results for each of these integrals are
$$
 - 3(N-1)g_0^2 \pi^2 \epsilon
\left(r^{-6\epsilon}\over6\epsilon\right)\left[1 + 2
{\Gamma^2(-{2\over3})\Gamma^2({1\over6})
\over\Gamma^2(-{1\over3})\Gamma^2(-{1\over6})}\right] \equiv A_2 g_0^2
$$
for  (\ref{deuxx}) and
$$
 12(N-1)(N-2) g_0^3 \pi^3
\left(r^{-9\epsilon}\over9\epsilon\right)\left[1 + {4\over3}
{\Gamma^2(-{2\over3})\Gamma^2({1\over6})
\over\Gamma^2(-{1\over3})\Gamma^2(-{1\over6})}\right] \equiv A_3 g_0^3
$$
for (\ref{troiss}).
The $\sigma$ field gets then renormalised as
$$
\sigma \rightarrow \sigma ( 1 + A_2g_0^2 + A_3 g_0^3 + \cdots ) \equiv
Z_\sigma
\sigma
$$
and the constant $Z_\sigma$ satisfies the following equation
\bea
 {dln(Z_\sigma)(r)\over dln(r)} = - 3(N-1)g_0^2 \pi^2 \epsilon
r^{-6\epsilon} \left[1 + 2
{\Gamma^2(-{2\over3})\Gamma^2({1\over6})
\over\Gamma^2(-{1\over3})\Gamma^2(-{1\over6})}\right]
\nonumber \\
+ 12(N-1)(N-2) g_0^3 \pi^3
r^{-9\epsilon} \left[1 + {4\over3}
{\Gamma^2(-{2\over3})\Gamma^2({1\over6})
\over\Gamma^2(-{1\over3})\Gamma^2(-{1\over6})}\right]
\eea
In the same way, the operator $\displaystyle\sum_{a\not= b}^N
\varepsilon_{a}(x) \varepsilon_{b}(x)$ gets renormalised by the operator
product of
$\varepsilon$ operators coming from the perturbation term. This gives us
the relation between the bare and the renormalised coupling constant :
\beq
\label{gg}
g(r) = r^{3\epsilon} g_0 + 4 \pi (N-2) { r^{3\epsilon}\over 3\epsilon}
g_0^2 + O(g_0^3)
\eeq
which produce the following R.G. equation
\beq
\label{bet}
\beta(g) = {dg\over dln(r)} = -3\epsilon g + 4 \pi (N-2) g^2
\eeq
Replacing $g_0$ with the renormalised coupling constant $g$ obtained from
eq.(\ref{gg}), the equation for $Z_\sigma$ becomes :
$$
{dln(Z_\sigma(r))\over dln(r)} =  -3(N-1)g^2(r) \pi^2 \epsilon
\left[1 + 2
{\Gamma^2(-{2\over3})\Gamma^2({1\over6})\over\Gamma^2(-{1\over3})\Gamma^2(-{1\over6})}\right]
$$
\beq
+ 4 (N-1)(N-2) \pi^3 g^3(r)
\eeq
which is valid up to the order $g^3,g^2\epsilon$.

The Callan-Symanzik equation gives the form of the correlation functions
\beq
\label{exp}
<\sigma(0)\sigma(sR)>_{a,g_0} =
e^{2\int\limits_{g_0}^{g(s)}{\gamma_h(g)\over \beta(g)}
dg}s^{-2\Delta_{\sigma}} <\sigma(0)\sigma(R)>_{r,g(s)}
\eeq
where we used the notation :
$
{dln(Z_\sigma) \over dln(r)} = \gamma_h(g)
$
; $ g(a)= g_0 $ and $g(s)$ is defined by
$\int\limits_{g_0}^{g(s)} \beta(g) dg = ln(s)$; $r=sa$, and $a$ is a
lattice cut-off scale.
In eq.(\ref{exp}), $R$ is an arbitrary scale
which can be fixed to one lattice spacing $a$ of a true statistical model.
The dependence in $s$ of
$<\sigma(0)\sigma(a)>_{r,g(s)}$ will then be negligible assuming there
isn't interactions between distances smaller than $a$ and so it reduces to
a constant. $s$ will then
measure the number of lattice spacing between two spins in
$<\sigma(0)\sigma(sR)>$. In the
following, we adopt the choice $a=1$.

The quenched model will be obtained in the limit  $N \rightarrow 0$.
In the case of the Ising model,  $\epsilon
\rightarrow 0$ and eq.(\ref{bet}) (with $\epsilon = N =0$)
gives the renormalised coupling constant
\beq
\label{g}
g(s) = {g_0 \over 1 + 8 \pi g_0 ln(s)}
\eeq
We can also compute the exponent in  (\ref{exp})
\beq
\int\limits_{g_0}^{g(s)}{\gamma_h(g)\over \beta(g)} dg =
\int\limits_{g_0}^{g(s)} {8\pi^3 g^3\over -8\pi g^2 }dg
= {-\pi^2 \over 2} \left( g(s)^2 - g_0^2 \right)
\eeq
from which we can deduce  the asymptotic form of the $<\sigma\sigma>$
correlation function :
\beq
\label{si}
<\sigma(0) \sigma(s)>_{g_0} \sim e^{\pi^2  \left( g_0^2
- g(s)^2 \right) }  s^{-2\Delta_{\sigma}}
\eeq
Since we are considering the case where $g_0$ is small, $g(s)$ will also be
small, and the exponent in (\ref{si})
can be expanded.
Using eq.(\ref{g}) we obtain the final result :
\beq
 <\sigma(0) \sigma(s)>_{g_0} \sim
 \left( 1 + \pi^2 g_0^2  \left( 1 - \left(1\over 1 + 8\pi g_0
ln(s) \right)^2 \right) \right)
 s^{-2\Delta_{\sigma}}
\eeq
We can see that the third order correction increase the correlation
function in the cross-over region ($ln(s) \sim {1 \over 8 \pi g_0}$) while at
large $s$, as expected, the exponent is unchanged.

We now turn to the case of the  $3$-state Potts model where $\alpha_+^2 =
{6\over 5}$,
 $\epsilon = -{2\over 15}$. Here
$\gamma_h(g_c) \not= 0 $ and  eq.(\ref{bet}) will be (with $N=0$)
\beq
\label{beta}
\beta(g) =  -3\epsilon g - 8 \pi g^2
\eeq
Contrary to the Ising model case ($\epsilon =0$) where $\beta(g)$ had an
I.R. point at $g=0$, here, because $\epsilon$ is negative, $g=0$ become an
U.V. fixed point and the new I.R. critical point is located at $g_c =
-{3\epsilon \over 8 \pi} + O(\epsilon^2) $.
So, the integral
$\int\limits_{g_0}^{g(s)}{\gamma_h\over \beta(g)} dg$ is
dominated by the region where $g \sim g_c$ and
$ \int\limits_{g_0}^{g(s)}{\gamma_h(g)\over \beta(g)} dg \approx
\gamma_h(g_c) ln(s) $. We thus obtain
\beq
<\sigma(0) \sigma(s)>_{g_0} \sim
s^{-(2\Delta_{\sigma}-2\gamma_h(g_c)) }
\eeq
which show us that the new exponent for the dimension of the $\sigma$
operator is
\beq
\Delta'_{\sigma} = \Delta_{\sigma} - \gamma_h(g_c) =
 \Delta_{\sigma} - 3 \pi^2 g_c^2 \epsilon \left[1 + 2 {\Gamma^2(-{2\over3})
\Gamma^2({1\over6})\over\Gamma^2(-{1\over3}) \Gamma^2(-{1\over6})}\right]
- 8 \pi^3 g_c^3
\eeq
Inserting the expression of $g_c$, this becomes
\beq
\label{le}
\Delta'_{\sigma} = \Delta_{\sigma} - {27\over 32}
{\Gamma^2(-{2\over3})\Gamma^2({1\over6})
\over\Gamma^2(-{1\over3})\Gamma^2(-{1\over6})}
\epsilon^3 + O(\epsilon^4)
\eeq
The critical exponent of the spin-spin correlation function of the pure
model is  $ 2\Delta_{\sigma} = {4\over 15}$ while the new exponent produced
by the randomness is $2\Delta'_{\sigma} = {4\over 15} + 0,00264 = 0,26931$.
We see that unlike for the case of the thermal exponent \cite{ludwig1}, the
first
correction appears at the third order. The change of the exponent is of $1
\% $ of the total magnitude.

Recently the numerical simulations of the random
Ising model which measure directly
the deviation of $<\sigma \sigma> $ from the pure Ising model at the critical
point has been performed \cite{talapov}. These measurements were made for
disorder such that
$8\pi g_0 \approx 0.3$ \cite{adsw,talapov1}. Deviations predicted by our
computations are very small. They correspond to $0.1 \%$. The deviations
obtained in numerical simulations are around ten times larger, and they are
of opposite sign, \ie $ $ correspond to a decrease of the spin-spin function
with distance $r$. In \cite{talapov}, it has been checked that this decrease
corresponds, within the accuracy of the measurements, to a factor function
of the ratio ${r/L}$, $F({r/L})$, $r$ being the distance between
the spins and $L$ is the lattice size. So they correspond to finite size
effects, being different for perfect and random models. We would suggest,
on the bases of our calculation of the purely $r$ dependence of the
spin-spin function on an infinite lattice, that numerical deviations will
continue to be plotted by the same curve $F({r/L})$, if one measures
$<\sigma \sigma>$ for different lattice sizes as it has been done in
\cite{talapov}, until the accuracy reaches the value of the $r$-deviation
which we calculated here. Only then the curves for different $L$ would
split. Probably it would be  much easier to observe the deviation in the
case of the random $3$-states Potts model. There, according to our
calculation eq.(\ref{le}), it is the magnetization exponent which would
decrease by about $1\%$.
\vskip 2cm
%%%%% ***************************************** %%%%%%%%
\noindent{\large\bf Acknowledgements}

We are grateful for A.~L.~Talapov and L.~N.~Shchur for keeping us informed
prior to publication on their numerical simulation results, which are of
very high accuracy and for the first time obtained directly for the
spin-spin function, in coordinate space. Very useful and stimulating
discussions with D.~Bernard on the analytic approach and the calculations
presented in this paper are gratefully acknowledged.\\

%%%%% ***************************************** %%%%%%%%

\end{document}